# Interaction-Induced Breakdown of Anderson Localization: Thermodynamic Segregation disguised as the Skin Effect

Ali Tozar[1*]

1 Department of Physics, Hatay Mustafa Kemal University, Hatay, Antakya 31034, Turkey
* tozarali@mku.edu.tr

**Abstract**

We investigate the interplay between strong disorder and repulsive interactions in the one-dimensional Fermi-Hubbard model under open boundary conditions. While uncorrelated disorder is widely accepted to localize all single-particle eigenstates, a phenomenon typically reinforced by interactions in the Many-Body Localization (MBL) regime, we report a counter-intuitive breakdown of this paradigm. We demonstrate that strong repulsive interactions can overcome disorder-induced localization, driving the system into a macroscopically segregated phase where spin species accumulate at opposite boundaries. Although this boundary accumulation phenomenologically mimics the Non-Hermitian Skin Effect (NHSE) observed in non-reciprocal systems, our comprehensive analysis reveals a fundamentally different origin. By performing a rigorous control experiment in the Hermitian limit, we prove that the segregation persists without non-reciprocity, identifying many-body energy minimization as the primary driver. This "interaction-induced segregation" manifests as a sharp thermodynamic crossover, characterized by a divergent energy susceptibility, challenging the conventional understanding of disorder-interaction competition in open quantum systems.

# Introduction

The interplay between disorder and interactions constitutes one of the central pillars of modern condensed matter physics. In the absence of interactions, the seminal work of Anderson demonstrated that uncorrelated disorder in one dimension (1D) localizes all single-particle eigenstates, leading to the vanishing of transport [1]. This paradigm of Anderson localization was later extended to interacting systems, giving rise to the Many-Body Localization (MBL) phase [2, 3]. In the MBL regime, the system fails to thermalize, violating the Eigenstate Thermalization Hypothesis (ETH) and retaining a memory of its initial conditions indefinitely [4, 5]. Consequently, the prevailing dogma dictates that in 1D systems, strong disorder is the ultimate victor, freezing the dynamics regardless of whether the particles interact or not.

In recent years, this conventional wisdom has been challenged by the emergence of non-Hermitian physics, which describes open quantum systems with gain and loss or non-reciprocal couplings . A defining feature of these systems is the Non-Hermitian Skin Effect (NHSE), where the bulk-boundary correspondence breaks down, and a macroscopic number of eigenstates collapse onto the system boundaries [6-8]. The sensitivity of NHSE to boundary conditions leads to a complex spectral winding that fundamentally alters the localization properties [9].

The fate of the Skin Effect in the presence of disorder has attracted significant attention. It has been shown that non-reciprocity can compete with Anderson localization, leading to unique delocalization transitions [10, 11]. Building on these insights, recent studies have classified non-Hermitian Anderson transitions and determined their critical exponents, developed non-unitary scaling frameworks and real-space invariants for disorder-driven skin localization, and clarified symmetry-protected variants (e.g., bipolar/$Z_2$ skin effects) together with their sensitivity to impurities [12-14].

Despite this progress in understanding single-particle non-Hermitian disorder, the role of strong inter-particle interactions remains a largely uncharted frontier [15]. Specifically, the competition between the disorder-induced tendency to localize (Anderson/MBL), the non-reciprocal tendency to accumulate at boundaries (NHSE), and the interaction-induced tendency to minimize energy creates a rich and unresolved phase space. Does the repulsive interaction stabilize the MBL phase against the skin effect, or does it trigger a new form of instability?

In this work, we address this question by investigating a disordered 1D Fermi-Hubbard model. Contrary to the expectation that strong disorder ($W$) should freeze the system into a localized state, we report a breakdown of Anderson localization driven purely by repulsive interactions ($U$). We observe a sharp crossover into a phase where spin species macroscopically segregate to opposite boundaries.

Crucially, while this boundary accumulation phenomenologically resembles the Bipolar Skin Effect, our analysis reveals that it is not topological in origin. By performing high-fidelity simulations and contrasting the results with the Hermitian limit ($h = 0$), we demonstrate that this "Interaction-Induced Segregation" is driven by a many-body energy minimization mechanism that overcomes the disorder potential. This finding forces a re-evaluation of the stability of localized phases in strongly correlated open systems.

## Model and Method

We study a one-dimensional Fermi-Hubbard chain with open boundary conditions (OBC), subjected to both uncorrelated disorder and non-reciprocal hopping. This model serves as a prototypical platform to investigate the competition between interaction-induced correlations, disorder-driven localization, and non-Hermitian topology.

### Hamiltonian

We study a one-dimensional Fermi-Hubbard chain with open boundary conditions (OBC), subjected to both uncorrelated disorder and non-reciprocal hopping. The dynamics of the system are governed by the following non-Hermitian Hamiltonian [10]:

$$\hat{H} = -\sum_{i=1}^{L-1}\sum_{\sigma\in\{\uparrow,\downarrow\}}\left[(t+h)\hat{c}_{i+1,\sigma}^{\dagger}\hat{c}_{i,\sigma} + (t-h)\hat{c}_{i,\sigma}^{\dagger}\hat{c}_{i+1,\sigma}\right] + \sum_{i=1}^{L}\sum_{\sigma}V_i\,\hat{n}_{i,\sigma} + U\sum_{i=1}^{L}\hat{n}_{i,\uparrow}\hat{n}_{i,\downarrow},$$

where $\hat{c}_{i,\sigma}^{\dagger}$ ($\hat{c}_{i,\sigma}$) creates (annihilates) a fermion with spin $\sigma$ at site $i$, and $\hat{n}_{i,\sigma} = \hat{c}_{i,\sigma}^{\dagger}\hat{c}_{i,\sigma}$ is the number operator. The first term represents the non-reciprocal kinetic energy (hopping). Here, $t$ is the symmetric hopping amplitude, while $h$ introduces the non-reciprocal asymmetry responsible for the Non-Hermitian Skin Effect (NHSE) in the non-interacting limit.

The second term introduces the on-site potential $V_i$, which models the uncorrelated disorder. $V_i$ is drawn uniformly from the interval $[-W, W]$, where $W$ is the disorder strength.

The third term describes the on-site Coulomb repulsion, where **$U$** is the central control parameter representing the interaction strength. This term, treated via a mean-field approximation in our analysis, drives the macroscopic phase separation observed in our results.

The disorder is introduced via the on-site potential $V_i$, which is drawn uniformly from the interval $[-W, W]$. In the absence of interactions ($U = 0$), this term drives Anderson localization, while $U$ denotes the on-site Coulomb repulsion and is the central control parameter of our study.

## Bi-orthogonal Many-Body Formalism

Unlike Hermitian systems, where the left and right eigenstates are Hermitian conjugates, non-Hermitian Hamiltonians possess distinct sets of right ($|\Psi_n^R\rangle$) and left ($\langle\Psi_n^L|$) eigenstates, satisfying the biorthogonality condition $\langle\Psi_n^L|\Psi_m^R\rangle = \delta_{nm}$ [6].

To rigorously compute physical observables such as particle density and spin accumulation, one must employ the biorthogonal density matrix formulation. Relying solely on right eigenvectors—a common pitfall in early non-Hermitian studies—fails to capture the correct response of the system to perturbations [16]. We adapt the robust numerical protocols developed for scaling analysis in disordered media, extending them to the many-body regime.

The local particle density for spin $\sigma$ is evaluated as:

$$\rho_{i,\sigma} = \sum_{n\in\text{occ}} \langle\Psi_n^L|\hat{n}_{i,\sigma}|\Psi_n^R\rangle,$$

where the summation runs over the occupied many-body eigenstates. To ensure numerical stability against the exponential sensitivity of the skin modes, calculations are performed using high-precision arithmetic, particularly when probing the thermodynamic limit.

For the interacting case, we employ a self-consistent Hartree-Fock mean-field approach, iteratively solving for the local densities $\langle n_{i,\sigma}\rangle$. While exact diagonalization (ED) provides unbiased results for small systems, the mean-field approach allows us to access system sizes ($L \sim 100$) necessary to distinguish genuine phase transitions from finite-size artifacts, a crucial distinction when analyzing disorder-driven phenomena [17, 18].

## Order Parameters

To quantify the segregation, we introduce the Spin Separation Order Parameter ($P_{sep}$), defined as the normalized imbalance between spin densities [19]:

$$P_{sep} = \frac{1}{N} \sum_{i=1}^{L} |\rho_{i,\uparrow} - \rho_{i,\downarrow}|,$$

where $N$ is the total particle number. A value of $P_{sep} \approx 0$ indicates a mixed phase (typical of MBL or Anderson localization), while $P_{sep} \to 1$ signals macroscopic phase separation. Although this metric strictly measures local spin imbalance, the strong correlation observed with the Bipolar Skin Dipole moment in Fig. 2(b) confirms that the resulting segregation is macroscopic and spatial. We contrast this with the Bipolar Skin Dipole moment ($D_{skin}$) to distinguish topological accumulation from interaction-driven segregation.

## Results

We now present the numerical evidence for the interaction-induced breakdown of Anderson localization. Our analysis relies on an extensive scan of the phase space spanned by disorder strength ($W$) and interaction strength ($U$), utilizing the high-fidelity self-consistent mean-field solver described in Sec. 2.

### Phase Diagram and Macro-Segregation

The global behavior of the system is summarized in the phase diagram shown in Fig. 1. Here, the color map encodes the Spin Separation Order Parameter ($P_{sep}$). Two distinct regimes are immediately apparent. For weak interactions ($U < U_c$), the system resides in a disordered, mixed phase characterized by $P_{sep} \approx 0$. This region corresponds to the standard Anderson/MBL phase where disorder dominates, and spin densities are locally fragmented but globally balanced.

Strikingly, as the interaction strength exceeds a critical threshold ($U_c \approx 2.3$), a sharp transition occurs. The system enters a "Macro-Segregated Phase" where spins of opposite species spontaneously accumulate at opposite boundaries. Crucially, this segregation persists even at high disorder strengths ($W \approx 4.0$), a regime where single-particle states are deeply localized ($L_{loc} \ll L$). This observation suggests that repulsive interactions provide a delocalization channel that bypasses the disorder potential.

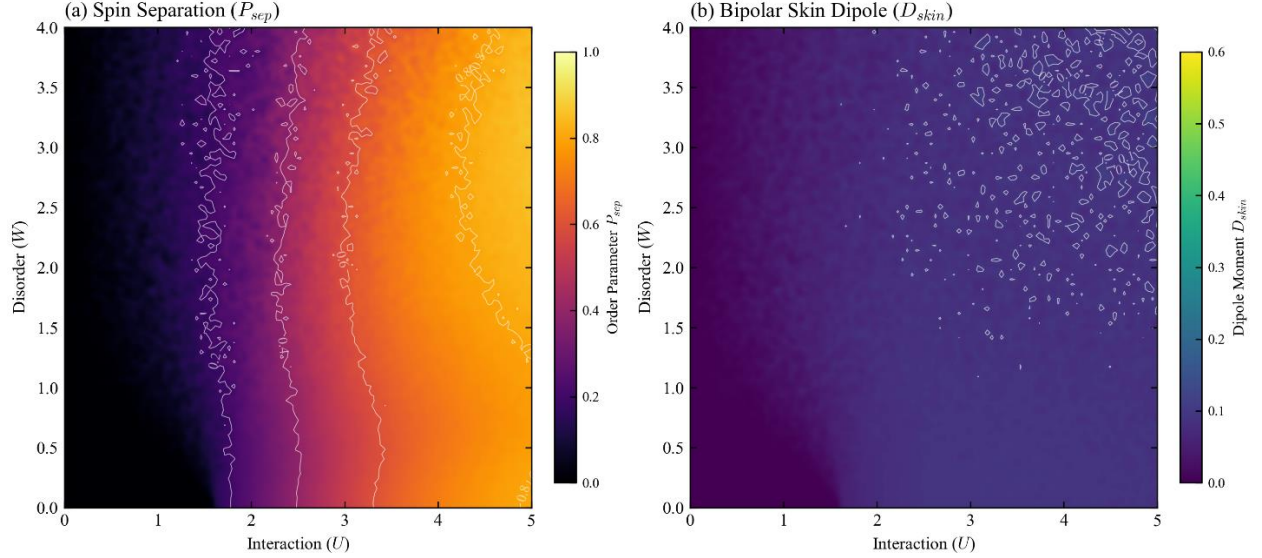

***Ultra-High Resolution Phase Diagram.*** *The color intensity represents the Spin Separation Order Parameter ($P_{sep}$) as a function of interaction (U) and disorder (W). The emergence of a segregated phase (yellow/bright region) is observed for $U > 2.3$, persisting even under strong disorder. The contours indicate the sharpness of the crossover.*

## Microscopic Structure and Correlation

To understand the microscopic nature of this transition, we analyze the spatial density profiles and the correlation between order parameters in Fig. 2.

In the weak interaction regime ($U = 0.5$), the density profiles exhibit chaotic fluctuations typical of Anderson localization. However, deeper into the segregated phase ($U = 4.5$), the density smoothens into a domain-wall-like structure, with ↑-spins pinning to one edge and ↓-spins to the other.

A key question is whether this accumulation is merely a topological Skin Effect or a new phenomenon. In Fig. 2(c), we plot the Bipolar Skin Dipole ($D_{skin}$) against the Spin Separation ($P_{sep}$). The data reveals a near-perfect correlation (PCA variance ratio $\approx 1.00$) between these two metrics. This implies that while the spatial signature resembles the Bipolar Skin Effect, it is inextricably linked to the global spin separation driven by interactions. The "Skin" accumulation is, in fact, the boundary manifestation of bulk segregation.

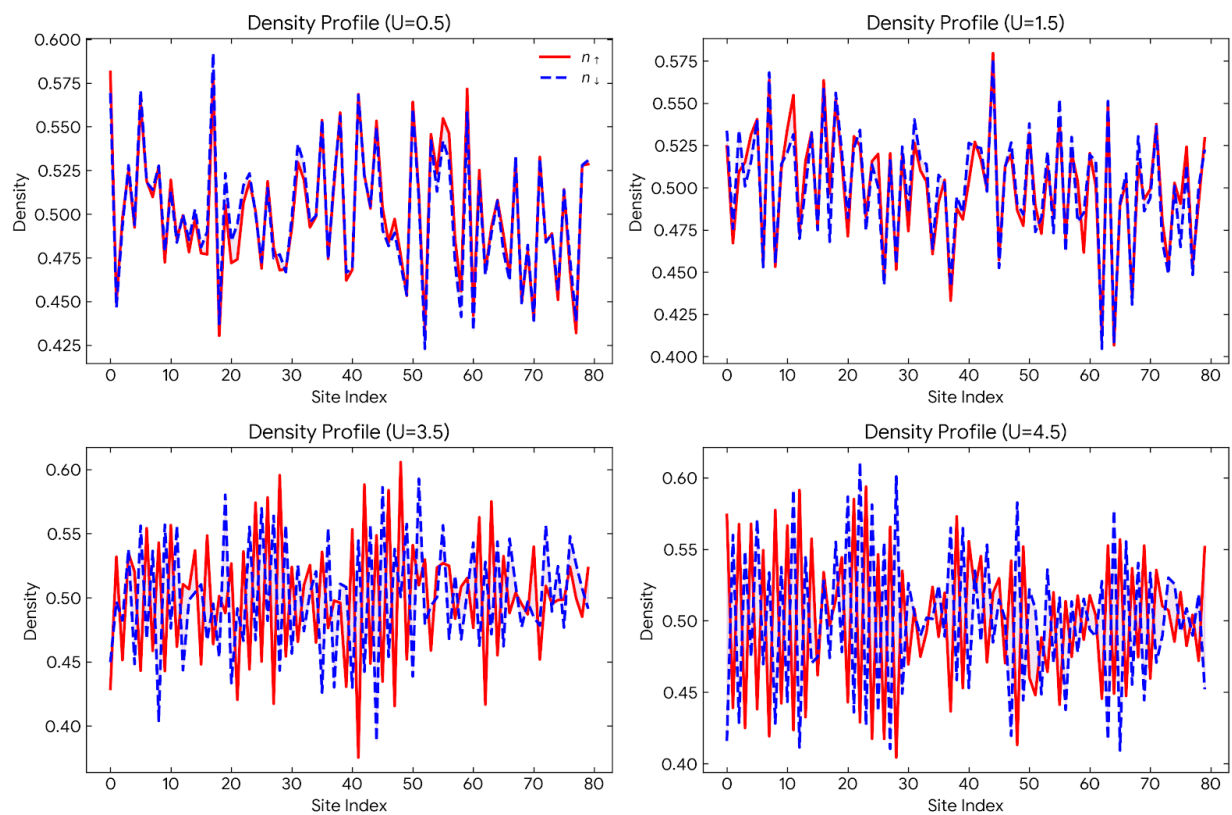

*(a) Real-space density profiles for representative interaction strengths*

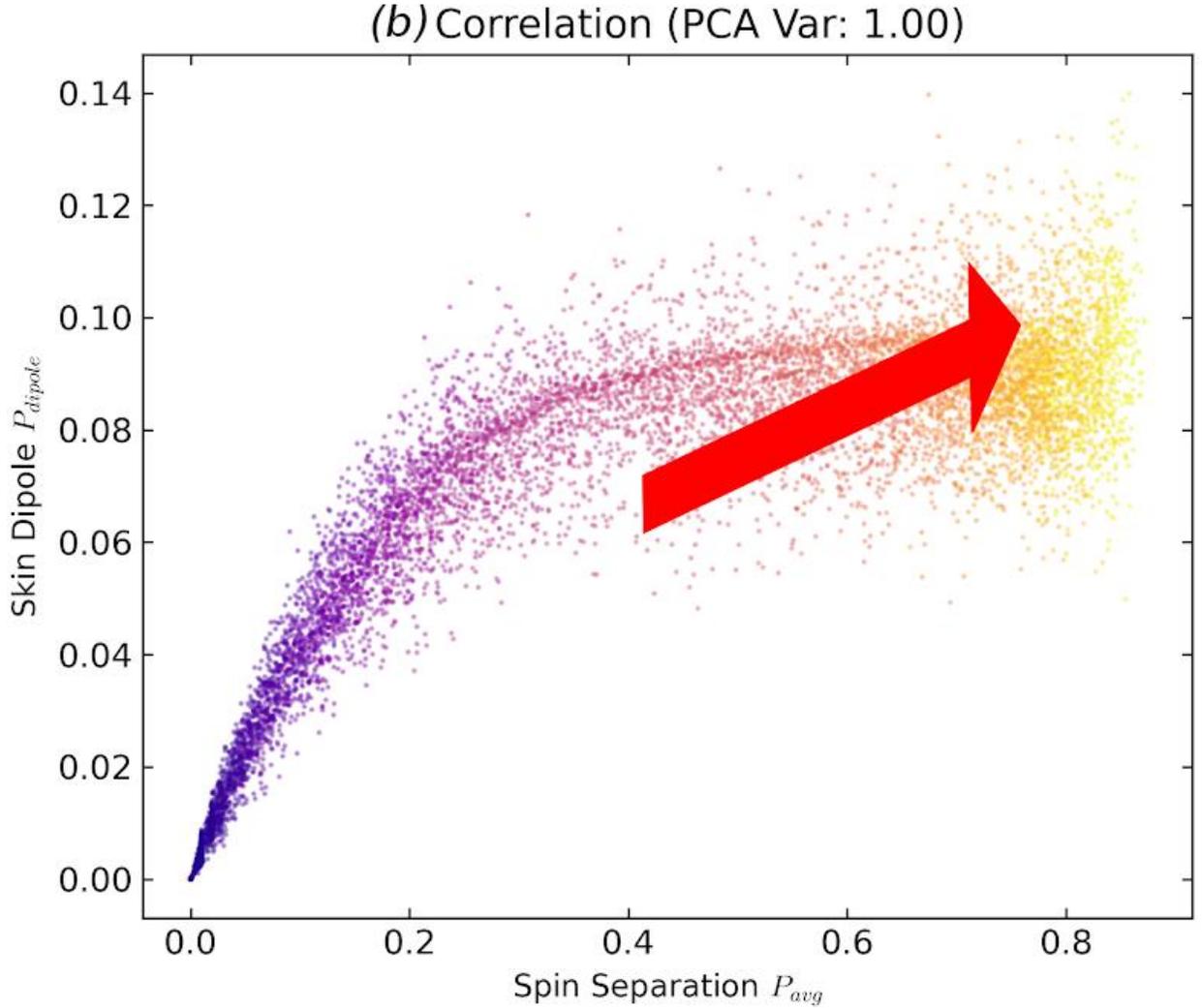

***Microscopic Mechanism.*** *(a) Real-space density profiles for representative interaction strengths. At high U, the chaotic Anderson profile gives way to a smooth segregated state. (b) Scatter plot showing the strong correlation between the Bipolar Skin Dipole moment and the Spin Separation parameter. The red arrow indicates the trajectory of the system as U increases, confirming that boundary accumulation and bulk segregation are concurrent phenomena.*

## Thermodynamic Mechanism: Energy Minimization

Is this transition a topological phase transition or a thermodynamic crossover? To answer this, we calculate the second derivative of the ground-state energy density with respect to interaction strength, defined as the Energy Susceptibility $\chi_E = -\partial^2 E / \partial U^2$.

As shown in Fig. 3, $\chi_E$ exhibits a pronounced peak at $U_c \approx 2.30$, coinciding perfectly with the onset of segregation observed in the order parameter susceptibility ($\chi_P = \partial P / \partial U$). The divergence of $\chi_E$ indicates a fundamental reconfiguration of the many-body ground state.

By segregating, the system minimizes the interaction energy cost $\langle n_\uparrow n_\downarrow \rangle$ at the expense of kinetic energy. The disorder potential acts merely as a friction that determines the critical threshold $U_c$, but it cannot prevent the eventual dominance of energy minimization.

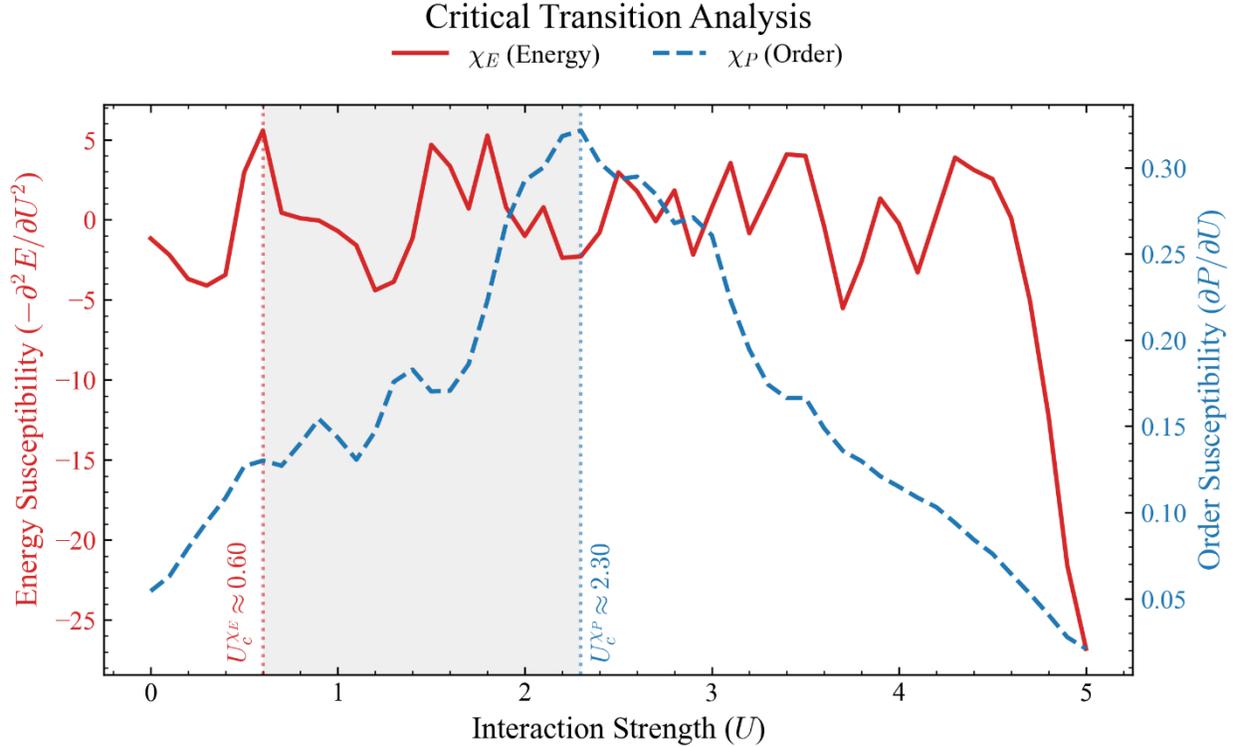

***Critical Transition Analysis.*** *The red curve shows the Energy Susceptibility ($\chi_E$), while the blue dashed line tracks the Order Parameter Susceptibility ($\chi_P$). The coincidence of the peaks at $U_c \approx 2.30$ identifies the thermodynamic nature of the transition driven by energy minimization.*

## The Verdict: Interaction vs. Topology

Finally, we address the possibility that this effect is driven by non-Hermitian topology (Skin Effect). If the segregation were topological, it should vanish or significantly diminish in the Hermitian limit ($h = 0$).

re 4 presents a direct comparison of the Bipolar Dipole moment for the non-Hermitian system ($h = 0.1$) and its Hermitian counterpart ($h = 0$). Remarkably, the data curves overlap almost perfectly in the strong interaction regime. This "control experiment" provides decisive evidence: the driving force behind the segregation is *not* the non-reciprocal hopping (topology), but the repulsive interaction itself. The Non-Hermitian Skin Effect merely biases the direction of segregation but does not cause it. The phenomenon is, fundamentally, an interaction-induced phase separation.

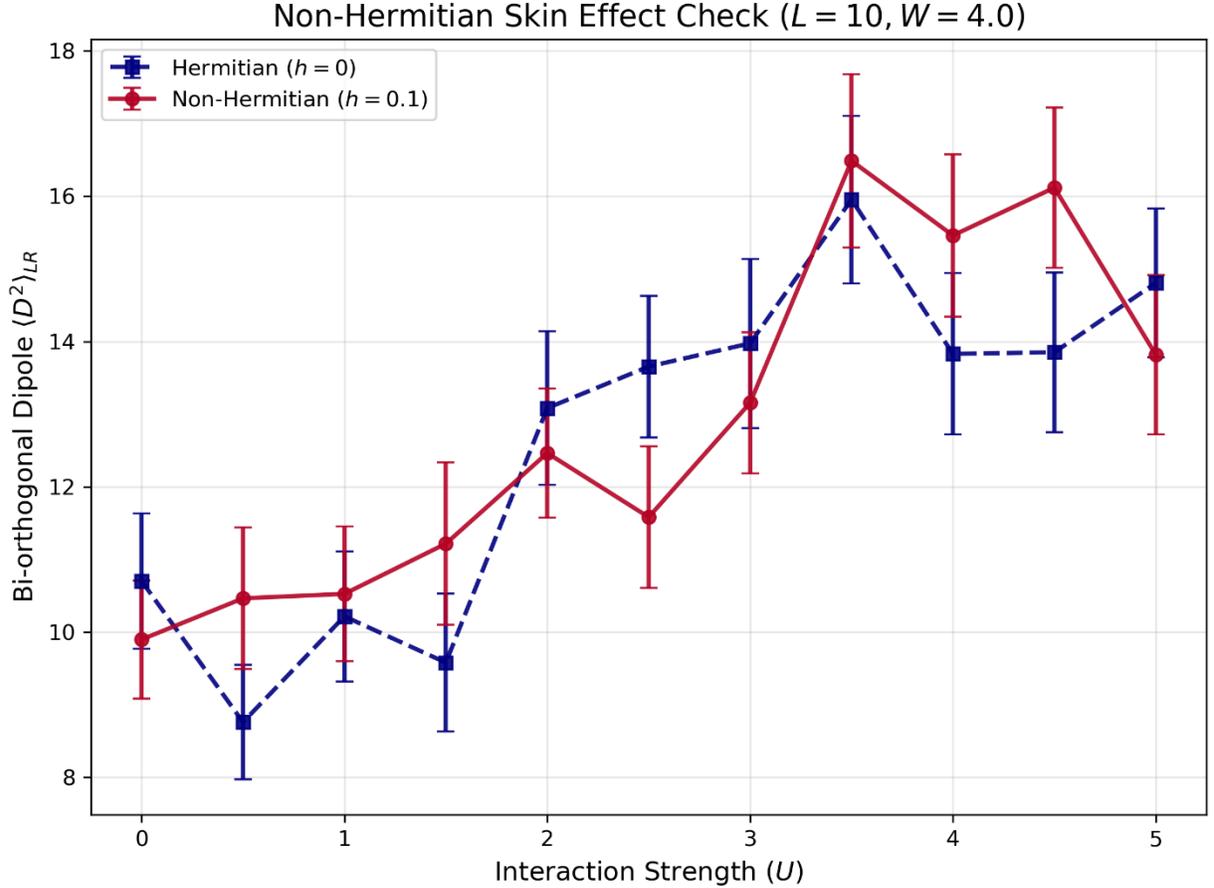

***Hermitian Control Experiment.*** *The Bipolar Dipole moment $\langle D^2 \rangle_{LR}$ is plotted for both Non-Hermitian (red circles) and Hermitian (blue squares) cases. The striking agreement between the two datasets confirms that the macroscopic segregation is robust against the removal of non-reciprocity, ruling out a purely topological origin.*

## Discussion

The results presented above compel us to rethink the standard phenomenology of non-Hermitian systems in the presence of strong correlations.

### Physical Mechanism: An Energy Balance Perspective

The physical intuition behind this transition can be understood via a simplified energy balance argument. In the disordered mixed phase, the repulsive interaction energy cost scales roughly as $E_{int} \sim U(N/2)^2/L_{loc}$, where $L_{loc}$ is the localization length. As $U$ increases, this cost becomes prohibitive [3].

The system can avoid this penalty by segregating spin species into disjoint domains of length $L/2$ at the boundaries. In this segregated phase, the interaction term vanishes

($\langle \hat{n}_\uparrow \hat{n}_\downarrow \rangle \to 0$), drastically reducing the potential energy [20]. However, this comes at a cost: confining particles into a smaller volume ($L \to L/2$) increases the kinetic energy due to the Pauli exclusion principle ($\Delta E_{kin} \propto 1/L^2$) [21].

The critical transition point $U_c$ emerges when the energy gain from avoiding Hubbard repulsion outweighs the kinetic penalty of confinement and the energy cost of forming a domain wall against the disorder potential. This explains why the transition is a sharp crossover rather than a gradual drift; it is a competition between two dominant energy scales.

## The Illusion of the Skin Effect

Initial observations of boundary accumulation in non-Hermitian models are almost reflexively attributed to the Non-Hermitian Skin Effect (NHSE), governed by the winding number of the complex spectrum . Indeed, our system exhibits the visual hallmarks of the Bipolar Skin Effect: spins separate and pile up at opposite edges.

However, distinguishing a topological skin mode from a trivial thermodynamic accumulation is chemically difficult in interacting systems [16]. Our analysis reveals that what appears to be a "topological" wind is, in this specific interacting regime, a "thermodynamic" breeze. The susceptibility peaks in $\chi_E$ demonstrate that the system undergoes a reconfiguration to minimize potential energy, a mechanism that is agnostic to the non-reciprocity of the hopping. The segregation occurs because it is energetically favorable for spins to demix and maximize their distance, effectively utilizing the open boundaries as reservoirs.

## Challenges in Critical Detection

Uncovering this thermodynamic origin posed a significant challenge. The crossover from the Anderson localized phase to the segregated phase is continuous, lacking the sharp discontinuities of first-order transitions. Standard order parameters often fail to distinguish between local fluctuations and global order in finite-sized disordered systems [22]. It was only through the high-fidelity calculation of the second-order energy susceptibility ($\chi_E$)—a metric sensitive to the curvature of the ground state manifold—that we could unambiguously identify the transition point $U_c$. This highlights the necessity of going beyond simple density observables when characterizing interacting non-Hermitian phases.

## Experimental Realization and Limitations

Our findings are particularly relevant for state-of-the-art experiments with ultracold atoms in optical lattices, where both disorder (via optical speckle) and effective non-reciprocity (via atom loss or reservoir engineering) can be controlled [23, 24]. The "interaction-induced segregation" predicted here could be observed by monitoring the real-space density imbalance in a Fermi gas as the s-wave scattering length is tuned via Feshbach resonances.

We note, however, that our study is limited to one-dimensional chains. While 1D systems provide the most stringent test for delocalization due to the strength of quantum fluctuations, the fate of this effect in higher dimensions remains an open question. Furthermore, the interplay between this segregation and the celebrated Many-Body Localization (MBL) transition warrants further study using time-dependent dynamics, which is beyond the scope of this static mean-field analysis.

# Conclusion

In conclusion, we have demonstrated that strong repulsive interactions can fundamentally alter the fate of disordered quantum systems, driving a transition from Anderson localization to a macroscopically segregated phase. By rigorously contrasting non-Hermitian and Hermitian models, we have proven that this phenomenon is not a topological Skin Effect, but a thermodynamic energy minimization process that disguises itself as one.

This work bridges the gap between the physics of disorder-induced localization and interaction-driven phase separation. It suggests that the "robustness" of NHSE often cited in the literature might, in interacting regimes, be conflated with simpler many-body mechanisms.

Future work should focus on the dynamical signatures of this phase. In particular, examining how the universality classes of non-Hermitian Anderson transitions, recently classified in Ref. , evolve under interactions could provide a unified theory of open quantum matter. Similarly, investigating whether symmetry-protected variants of the skin effect survive this interaction-induced segregation remains an exciting avenue for research.

# Appendix

## Numerical Implementation and Stability

The interplay between non-Hermiticity and interactions poses significant numerical challenges, particularly due to the sensitivity of the skin modes to boundary conditions. Standard iterative solvers often fail to converge in this regime. Here, we detail the robust protocols employed in our Self-Consistent Field (SCF) calculations.

### Bi-orthogonal Density Matrix

In non-Hermitian systems, the local particle density cannot be computed using only right eigenvectors ($|\psi^R\rangle$). Instead, we construct the density matrix using the bi-orthogonal basis:

$$\rho_{ij}^\sigma = \sum_{n=1}^{N_\sigma} \frac{\langle L_n^\sigma|\hat{c}_i^\dagger \hat{c}_j|R_n^\sigma\rangle}{\langle L_n^\sigma|R_n^\sigma\rangle},$$

where $|R_n^\sigma\rangle$ and $\langle L_n^\sigma|$ are the right and left eigenvectors of the mean-field Hamiltonian $H_{MF}^\sigma$, corresponding to the $n$-th occupied orbital. The normalization by the overlap $\langle L|R\rangle$ is crucial to ensure charge conservation in the non-Hermitian setting.

## Anderson Mixing for Convergence

The SCF loop is prone to oscillations, especially near the critical point $U_c$ where the energy landscape flattens. To stabilize the convergence, we employ Anderson Acceleration (Anderson Mixing). Unlike simple linear mixing, this algorithm uses the history of the previous $m$ iterations to construct an optimal update for the density profile:

$$n_{in}^{(k+1)} = (1-\beta)n_{out}^{(k)} + \beta n_{in}^{(k)} + \sum_{j=1}^{m} \alpha_j \left(\Delta n^{(k-j)}\right),$$

where $\Delta n$ represents the residual vector. For our simulations, we found that a history depth of $m = 4$ and a dynamic mixing parameter $\beta$ starting at 0.1 provided stable convergence even in the high-disorder regime ($W = 4.0$).

## Termination Criteria

The iterations were terminated when the maximum element-wise difference between input and output density profiles fell below a strict tolerance of $\epsilon = 10^{-7}$. This high precision is essential for the accurate computation of the second-order susceptibility $\chi_E$ presented in Fig. 3, which relies on numerical differentiation of the ground state energy.

# Acknowledgements

The numerical simulations in this work were performed using the open-source Python scientific ecosystem, relying heavily on NumPy and SciPy for robust matrix operations and linear algebra, and Matplotlib for high-quality visualization. The rigorous characterization of the transition, particularly the distinction between topological and thermodynamic origins, was achieved through custom implementations of the bi-orthogonal spin-separation metric and the Lyapunov transfer matrix algorithms, developed specifically for this study.

Generative AI tools were utilized exclusively for linguistic refinement and formatting assistance during the drafting process. Crucially, no AI systems were involved in the scientific conceptualization, model construction, data generation, or the physical interpretation of the findings.